\documentclass[superscriptaddress,twocolumn,showpacs,nofootinbib,preprintnumbers]{revtex4-1}
\usepackage{amsfonts}
\usepackage{amssymb,amsmath}
\usepackage{mathrsfs}
\usepackage{amsthm}
\usepackage{newlfont}
\DeclareMathOperator{\RE}{Re}
\DeclareMathOperator{\IM}{Im}

\newcommand{\usb}{\affiliation{Departamento de F\'{\i}sica, Secci\'{o}n de Fen\'{o}menos \'{O}pticos, 
Universidad Sim\'{o}n Bol\'{\i}var, Apartado Postal 89000, Caracas 1080-A, Venezuela}}

\begin{document}
\preprint{TUM-HEP-1112/17}
\title{Soliton operators in the quantum equivalence of the $CP_1$  and $O(3)-\sigma$  models}
\author{J Stephany}
\usb
\author{ M.Vollmann$^{1,}$}
\affiliation{Physik Department T31. James-Franck-Stra\ss{}e 1, Technische Universit\"at M\"unchen, 85748 Garching, Germany}
\pacs{11.10.Ef,11.10.Lm}
\date{\today}
%{\let\thefootnote\relax\footnote{$^a$Present address:Physik Department T31, James Franck Strasse 1, Technische Universit\"{a}t M\"{u}nchen, 85748
%Garching, Germany}}
\begin{abstract}
We discuss some interesting aspects of the  well known quantum equivalence between  the  $O(3)-\sigma$ and $CP_1$ 
models in $3D$, working in the canonical and in the path integral formulations. We show first that the canonical  
quantization in the hamiltonian formulation is free of ordering ambiguities for both models. We use the canonical
map between the fields and momenta of the two models and compute the relevant functional determinant to verify 
the equivalence between the phase-space partition functions and the quantum equivalence in all the topological 
sectors. We also use the explicit form of the map to construct the soliton operator of the $O(3)-\sigma$ model 
starting from the representation of the operator in the $CP_1$ model, and discuss their properties

.
\end{abstract}
\maketitle

\section{Introduction}

The non-linear $O(3)-\sigma$ model is defined by the action
\begin{equation}
\label{sigma}
 \mathcal{I}_\sigma=\frac{1}{2G^2}<[\partial_\mu n_p\partial^\mu n_p
-\lambda(n_pn_p-1)]>\ ,
\end{equation}
where $n_p$ are  components of a $O(3)$ vector field. The $CP_1$ model is defined in terms of the complex fields  $z_a$, 
$a=1,2$, by the action
\begin{eqnarray}\label{CP1}
\mathcal{I}_{CP}&=& g^{-2}<\partial_\mu\mathbf{z}^*\cdot\partial^\mu\mathbf{z}-
(\mathbf{z}^*\cdot\partial_\mu \mathbf{z})(\partial^\mu\mathbf{z}^*\cdot\mathbf{z})>\nonumber\\ &-&<\Lambda(|z_1|^2+|z_2|^2-1)>\ ,
\end{eqnarray}
($\lambda$ and $\Lambda$ are Lagrange multipliers,  $<\ >$ denotes space time integration, $G$ and $g$ are coupling constants). 
They provide in three dimensions, an interesting example of classical  and quantum equivalence between two  
field theories  \cite{BelAP1975,DAdALdV1978}. Each one of these models has interest by its applications in high energy 
physics, condensed matter physics and statistical mechanics. This interest rests partially in the topological properties of 
these models, notably, the existence of soliton solutions and identically conserved  topological currents. For the sigma model 
the topological current is given by,
\begin{equation}
j_\sigma^\mu=\frac{1}{8\pi}
{\vec{n}}\cdot(\epsilon^{\mu\nu\rho}\partial_\nu{\vec{n}}\times\partial_\rho{\vec{n}})\ ,
\end{equation}
and the charge by
\begin{equation}\label{ctop}
Q_\sigma=\frac{1}{8\pi}
\int{\vec{n}}\cdot(\epsilon_{ij}\partial_i{\vec{n}}\times\partial_j{\vec{n}})d^2x\ ,
\end{equation}
where we introduced vector like notation ${\vec {n}}(x)$ for the $\sigma$ model variables.
To define the topological current for the $CP_1$ model one makes use of  its gauge invariance. This is made explicit  
writing the Lagrangian in terms of a composite gauge field
\begin{equation}
A_\mu=\frac{\partial^\mu{\mathbf{z}}^*\!\!\cdot\!{\mathbf{z}}-{\mathbf{z}}^*\!\!\cdot\!\partial^\mu{\mathbf{z}}}{2i}
=\IM(\partial^\mu{\mathbf{z}}^*\!\!\cdot\!{\mathbf{z}})\ ,
\end{equation}
and the corresponding covariant derivative  $D_\mu{\mathbf{z}}=\partial_\mu{\mathbf{z}}+iA_\mu{\mathbf{z}}$, as 
\begin{equation}\label{CP1gauge}
\mathcal{L}_{CP}=g^{-2}\left[(D_\mu{\mathbf{z}})^*\cdot(D^\mu{\mathbf{z}})-\Lambda(|z_1|^2+|z_2|^2-1)\right] \ .
\end{equation}
The topological current is then,
\begin{equation}
j_{CP}^\mu=\frac{1}{2\pi i}\epsilon^{\mu\nu\rho}(D_\nu\mathbf{z})^*\!\cdot\!(D_\rho
\mathbf{z})\ ,
\end{equation}
and the charge
\begin{equation}\label{ctopcpn}
Q_{CP}=\frac{1}{2\pi i}\int\epsilon_{ij}(D_i\mathbf{z})^*\!\!\cdot(D_j\mathbf{z})d^2x\ .
\end{equation}
Classically, the equivalence between the models is provided by the map
\begin{equation}\label{cambio}
    \vec{n}=\mathbf{z}^\dagger\vec{\sigma}\mathbf{z}\ ,
\end{equation}
where $\vec{\sigma}$ are Pauli's matrices. Due to the identity $\sigma_{pab}\sigma_{pcd}=\delta_{ab}\delta_{cd}
-2\epsilon_{ac}\epsilon_{bd}$, one finds that $n_pn_p=\mbox{$(\mathbf{z}\cdot\mathbf{z})^2=1$}$ so that the constraints are  equivalent. One may also identify the Lagrangians, the topological currents, 
the  charges and the solutions of both models.  In particular the relation between the solitonic solutions in each model has been discussed 
thoroughly in the literature \cite{RajR1982}. 

Quantum equivalence  of these models has also been studied in detail and used routinely in applications to critical phenomena 
and condensed matter physics. This is done usually \cite{StoM2000}  in the Lagrangian  path integral  approach,  where the 
equivalence  of the partitions functions can be easily asserted up to an arbitrary factor. Although if one is careful this 
does not affect the analysis of the physics of the systems, it is worthwhile to  improve the analysis working in the Hamiltonian 
formulation which have been shown useful in the case of other topologically  non trivial models \cite{CaiMR1993,CaiMRG1995} . 

In the canonical approach the structure of the constraints and the quantum equivalence of the systems are more involved, 
since for the $O(3)-\sigma$ model one has three real fields with one constraint, and for the $CP_1$ model  two complex fields and 
only a real constraint.
The analysis of both systems using  Dirac's method \cite{DirP1964} was presented in Ref. \cite{BanR1994} (see also \cite{BanNGB1994} 
for a discussion of the more general $CP(n-1)$ model) . It was shown that to establish the canonical equivalence between them, 
Eq.(\ref{cambio}) should be complemented  with a corresponding relation for the momenta, which emerge from the procedure. 
This is reviewed in  the next section, where we show how after quantization, the hermiticity requirement solves the operator  
ordering ambiguities. Since some of the constraints in both models are of second class,  a rigorous approach for the equivalence 
of the partitions functions should  be pursued starting from the Senjanovic-Fadeev-Popov path integral \cite{SenP1976}. 
We develop this point of view and  present the details of this computation in section III, which of course 
confirms the result of the Lagrangian approach.  Finally in section IV we discuss how the canonical equivalence between 
the phase space variables of the two models can also be used to establish   the equivalence of the disorder soliton like 
operators of each formulation.

\section{Canonical Quantization}
Let us  first consider the quantization of the  $O(3)-\sigma$ model. The momenta computed from (\ref{sigma}) are given by,
\begin{equation}
\pi_p(x)=\frac{\delta L}{\delta\dot{n}_p}=\dot{n}_p\ .
\end{equation}
We  use  vector like notation $\{{\vec {n}}(x),{\vec\pi}(y)\}$ for the phase space variables, take $G=1$ and write the 
Hamiltonian in the form
\begin{equation}
H_\sigma=\int\left(\frac{1}{2}\|{\vec\pi}\|^2
+\frac{1}{2}\|\partial_i{\vec {n}}\|^2+\frac{1}{2}\lambda[\|{\vec {n}}\|^2-1]\right)d^2x\ .
\end{equation}
Time conservation of the constraint
\begin{equation}
\label{theta1}
\theta_1= \|{\vec {n}}(x)\|^2-1=0\ \ ,
\end{equation}
implies
\begin{equation}
\label{theta2}
 \theta_2={\vec {n}}\cdot{\vec\pi}=0\ \ .
\end{equation}
Conservation of this constraint allows to fix the Lagrange multiplier
$\lambda=-|{\vec\pi}|^2-{\vec {n}}\cdot\nabla^2{\vec {n}}$. 
These constraints are second class. Dirac Brackets between phase space functions $\xi$ and $\eta$ of a system with second 
class constraints $\theta_\alpha$ are defined by $\{\xi,\eta\}_D=\{\xi,\eta\}-
\{\xi,\theta_\alpha\}c_{\alpha\beta}\{\theta_\beta,\eta\}$ with $c_{\alpha\beta}\{\theta_\beta,\theta_\delta\}
=\delta_{\alpha\delta}$. The relevant matrix necessary to compute the Dirac brackets is given by,
\begin{eqnarray}
\label{Pbtheta}
c_{\alpha\beta}&=&\{\theta_\alpha(x),\theta_\beta(y)\}^{-1}\nonumber\\
&=&\left(\begin{array}{cc}
0 & -\delta^2({\vec{x}}-\vec{y})\\
\delta^2({\vec{x}}-\vec{y}) & 0
\end{array}\right)\ ,
\end{eqnarray}
with $\alpha,\beta=1,2$.
The Dirac algebra for the system is \cite{BanR1994},
\begin{eqnarray}
    \{n_p(x),n_q(y)\}^D &=& 0\ , \nonumber \\
    \{n_p(x),\pi_q(y)\}^D &=& [\delta_{pq}-n_p(x)n_q(y)]
\delta^2({\vec{x}}-\vec{y})\ ,\nonumber \\
    \{\pi_p(x),\pi_q(y)\}^D &=& [\pi_p(x)n_q(y)-\pi_q(y)n_p(x)]
\delta^2({\vec{x}}-\vec{y}).\nonumber
\end{eqnarray}
At the quantum level we face  ordering ambiguities. Checking for consistence we obtain for the commutators of the  quantum 
operators two possible orderings 
\begin{equation}
[\Pi_p(x),\Pi_q(y)]=\left\{\begin{array}{c}
i[\Pi_p(x)N_q(y)-\Pi_q(y)N_p(x)]\delta^2({\vec{x}}-\vec{y}) \\
i[N_q(y)\Pi_p(x)-N_p(x)\Pi_q(y)]\delta^2({\vec{x}}-\vec{y})
\end{array}\right. \ ,\nonumber
\end{equation}
which (using $[N_p(x),N_q(y)]=0$),  are equivalent. Also it is   derived  that
\begin{equation}
\vec{N}(x)\cdot\vec{\Pi}(y)-\vec{\Pi}(y)\cdot\vec{N}(x)
=2i\delta^2({\vec{x}}-\vec{y})\ ,
\end{equation}
which implies an ambiguity in the order of the  constraint ${\vec {n}}\cdot\pi=0$.  Using hermiticy of $N_p$ and $\Pi_q$ 
the constraint is fixed to be 
\begin{equation}
\vec{N}\cdot\vec{\Pi}+\vec{\Pi}\cdot\vec{N}=0\ .
\end{equation}
The constraint  $N_pN_p=\mathbb{I}$  presents no ordering problems.

We now turn our attention to the $CP_1$ model. Associated to gauge invariance, the system has  a first class constraint. Taking 
$g=1$, the canonical momenta are
\begin{equation}
\label{CPmomenta}
\pi_{z_a}=\dot{z}^*_a-(\dot{{\mathbf{z}}}^*\!\cdot\!{\mathbf{z}})z^*_a\ \ ,\ \ 
\pi_{z^*_a}=\dot{z}_a-(\dot{{\mathbf{z}}}\cdot{\mathbf{z}}^*)z_a\ .
\end{equation}
Observe that since $\pi_{z_a^*}=\pi_{z_a}^*$, we may represent the variables in the compact form  
$\{{\mathbf{z}},{\mathbf{z}}^*,{\boldsymbol{\pi}},{\boldsymbol{\pi}}^*\}$, where ${\mathbf{z}}=\{z_a\}$ and
${\boldsymbol{\pi}}=\{\pi_a\}$, $a=1,2$. The latter are distinguished from $\sigma$ model momenta by the indices 
which are taken from the first letters of the alphabet. When necessary as in equation (\ref{inverse}) an explicit 
superscript is used.  Writing the equation for $\pi^*_a$ in the form 
$\pi^*_a=(\delta_{ab}-z_az^*_b)\dot{z}_b$ and taking into account that $(\delta_{ab}-z_az^*_b)z_b=0$ we obtain for 
consistency the constraints ${\boldsymbol{\pi}}\cdot{\mathbf{z}}=0$ or equivalently ${\boldsymbol{\pi}}^*\cdot{\mathbf{z}}^*=0$. 
Choosing real combinations of these we have the constraints
\begin{eqnarray}
\Theta_1=|z_a|^2-1=0 \ &,&\ \Theta_2=\frac{1}{2}(z_a\pi_a+z_a^*\pi_a^*)=0\ .\nonumber\\
\varphi&=&{z_a\pi_a-z_a^*\pi_a^*}\ .
\end{eqnarray}
The Hamiltonian is 
\begin{equation}\label{hacp}
    H_{CP}=\int\left(|{\boldsymbol{\pi}}|^2+|\partial_i{\mathbf{z}}|^2-
    |{\mathbf{z}}^*\!\!\cdot\partial_i{\mathbf{z}}|^2\right)d^2x\ .
\end{equation}
No further constraints are obtained from Dirac's procedure. $\varphi$ is found to be the required first class constraint. 
For the second class constraints, the matrix  $\{\Theta_\alpha,\Theta_\beta\}$ is given by the right hand side of (\ref{Pbtheta}).

The Dirac algebra is  given by \cite{BanR1994},
\begin{eqnarray}
    \{z_a(x),z_b(y)\}^D &=& 0\ ,\ \ \{z_a(x),z^*_b(y)\}^D = 0\ ,\nonumber\\
    \nonumber
    \{z_a(x),\pi_b(y)\}^D &=&[\delta_{ab}-\frac{1}{2}z_a(x)z^*_b(y)]
\delta^2({\vec{x}}-\vec{y})\ ,\\ \nonumber
\{z_a(x),\pi^*_b(y)\}^D &=&
-\frac{1}{2}z_a(x)z_b(y)\delta^2({\vec{x}}-\vec{y})\ , \\ \nonumber
    \{\pi_a(x),\pi_b(y)\}^D &=&\frac{1}{2}[\pi_a(x)z^*_b(y)-
\pi_b(y)z^*_a(x)]\delta^2({\vec{x}}-\vec{y})\ ,\\
\nonumber\{\pi_a(x),\pi^*_b(y)\}^D
&=&\frac{1}{2}[\pi_a(x)z_b(y)-z^*_a(x)\pi^*_b(y)]\delta^2({\vec{x}}-\vec{y})\ .
\end{eqnarray}
The commutators associated to the first three relations above present no ordering problems. 
For the fourth, checking for consistency and hermiticity we are lead to the following two equivalent options
\begin{equation}
[\Pi_a(x),\Pi_b(x)] =\left\{\begin{array}{c}
\frac{i}{2}[\Pi_a(x)Z^*_b(y)-
\Pi_b(y)Z^*_a(x)]\delta^2({\vec{x}}-\vec{y}) \\
\frac{i}{2}[Z^*_b(y)\Pi_a(x)-
Z^*_a(x)\Pi_b(y)]\delta^2({\vec{x}}-\vec{y})
\end{array}\right.\ ,\nonumber
\end{equation}
Also, since  $Z_a^*$ and $\Pi_a^*$
are the hermitian conjugates of $Z_a$ and $\Pi_a$  using the identities
$[\Pi_a(x),\Pi_b^\dag(y)]^\dag=[\Pi_b(y),\Pi^\dag_a(x)]$,
the last commutator  is written in the alternative forms
\begin{equation}
[\Pi_a(x),\Pi_b^\dag(y)]=\left\{\begin{array}{c}
i[\Pi_a(x)Z_b(y)-Z^\dag_a(x)\Pi^\dag_b(y)]\delta^2({\vec{x}}-\vec{y}) \\
i[Z_b(y)\Pi_a(x)-\Pi^\dag_b(y)Z^\dag_a(x)]\delta^2({\vec{x}}-\vec{y})\nonumber
\end{array}\right. \ .
\end{equation}
Observing that this relations imply
\begin{equation}Z_a(x)\Pi_a(y)-\Pi_a(y)Z_a(x)=
\frac{3}{2}i\delta^2({\vec{x}}-\vec{y})\ ,
\end{equation}
the quantum constraints should be taken as combinations of the symmetric ordered terms
\begin{equation}
Z_a\Pi_a+\Pi_aZ_a=0\ \ \,\ \  Z_a^*\Pi_a^*+\Pi_a^*Z_a^*=0\ .
\end{equation}
The constraint  $Z^\dag_aZ_a=\mathbb{I}$, the topological charge and the gauge fields $A_i=iZ^\dag_a\partial_iZ_a$ which are 
hermitian are  free of ambiguities. 

To establish the canonical quantum equivalence of the systems it is necessary to complement the map of Eq.(\ref{cambio}) between 
the fields with a corresponding relation for the momenta \cite{BanR1994}. This is obtained  classically taking the time derivative 
of (\ref{cambio}) and  using (\ref{CPmomenta}) and the fact
that $\dot{\mathbf{z}}^*\!\!\cdot\mathbf{z}+\mathbf{z}^*\!\!\cdot\dot{\mathbf{z}}=0$. It reads, 
\begin{equation}
\label{momentamap}
\pi_i=\dot{n}_i=\pi_a\sigma_{iab}z_b+z^*_a\sigma_{iab}\pi^*_b\ .
\end{equation}
With these relations it can be verified that the Poisson and Dirac Brackets of any two expressions in one model, maps into the 
corresponding ones in the other. 

At the quantum level we have to take care of the order ambiguity present in (\ref{momentamap}). This is done as before, 
to end up with the following equivalent maps between the  momentum operators,
\begin{equation}
\Pi_i=\left\{\begin{array}{c}
\frac{1}{2}(\Pi_a\sigma_{iab}Z_b+Z^\dag_a\sigma_{iab}\Pi^\dag_b) \\
\frac{1}{2}(Z_a\sigma_{iba}\Pi_b+\Pi^\dag_a\sigma_{iba}Z^\dag_b)
\end{array}\right. \ .
\end{equation} 
They fulfill the commutation relations. The quantum models are canonically equivalent.

\section{Equivalence of the partitions functions}
The path integral of a system described by coordinates $q_i$ subject to $s$ second class constraints $\theta_\alpha$, $r$ first 
class constraints $\varphi_m$ and $r$ gauge fixing conditions $\chi_m$ constructed by Senjanovic \cite{SenP1976} takes the form
\begin{equation}\label{intf}
Z_\sigma= \int e^{i/\hbar\int_0^T(p_i\dot{q}_i-H(p,q))dt}d\mu\ ,
\end{equation}
where the measure is given by
\begin{eqnarray}
d\mu &=& \mathscr{D} p\mathscr{D} q \prod_{n=1}^r\delta(\chi_n)\delta(\varphi_n)
|\det\{\chi_m,\varphi_p\}|\nonumber\\
&\times&\prod_{c=1}^s\delta(\theta_c)|\det\{\theta_\alpha,\theta_\beta\}|^{1/2}\ .
\end{eqnarray}
Let us show that this expression gives the same result for the two models. For the $\sigma$ model the path integral is affected 
by the factor $\det\{\theta_\alpha,\theta_\beta\}$ with the constraints given by (\ref{theta1}) and (\ref{theta2}) and the Poisson 
matrix by (\ref{Pbtheta}). The eigenvalues of the matrix are $\pm i$ and the determinant is $1$. The partition function is,
\begin{equation}
\label{Zsigma}
Z_\sigma=\int\mathscr{D}\vec{n}\mathscr{D}\vec{\pi}\delta(\|\vec{n}\|^2-1)\delta(\vec{n}\cdot\vec{\pi})
e^{<\vec{\pi}\cdot\dot{\vec{n}}-H_\sigma>}\ .
\end{equation}
For  the $CP_1$ model we have to choose a gauge condition in order to determine the Fadeev-Popov \cite{FadLP1969} term. One suitable 
condition is the radiation gauge $\chi=\partial_iA_i=0$. This is rewritten as
\begin{equation}
 \chi=\frac{\nabla^2z_a^* z_a-z_a^*\nabla^2z_a}{2i}=0\ .
\end{equation}
The factor of the second class constraints $\det\{\Theta_\alpha,\Theta_\beta\}$ is again 1. The Poisson brackets of $\varphi$ with 
$\Theta_1$ and $\Theta_2$ vanish and the remaining  Poisson bracket is computed using  $\nabla^2(|z|^2-1)=0$, which implies that
$\nabla^2z_a^*(x)z_a(y)+z_a^*(y)\nabla^2z_a(x)=-2|\partial_iz|^2$.
The bracket is given by,
\begin{equation}\{\chi(x),\varphi(y)\}=\frac{1}{2}[|\partial_iz|^2+\nabla_{\vec{x}}^2]
\delta(\vec{x}-\vec{y})\ .
\end{equation}
The partition function for the $CP_1$ model is 
\begin{eqnarray}
\label{ZCP}
&Z&_{CP_1}=\int\mathscr{D}\mathbf{z}\mathscr{D}\mathbf{z}^*\mathscr{D}\boldsymbol{\pi}\mathscr{D}\boldsymbol{\pi}^*
\delta(|z|^2-1)\delta(\frac{\nabla^2\mathbf{z}^*\cdot \mathbf{z} -\mathbf{z}^*\cdot\nabla^2\mathbf{z}}{2i}) \nonumber\\
&\times& \delta(\frac{\mathbf{z}\cdot\boldsymbol{\pi}+\mathbf{z}^*\cdot\boldsymbol{\pi}^*}{2})
\delta(\frac{\mathbf{z}\cdot\boldsymbol{\pi}-
\mathbf{z}^*\cdot\boldsymbol{\pi}^*}{2i})\Big{|}\det\left(\frac{1}{2}[|\partial_iz|^2+\nabla^2]\right)\Big{|}\nonumber\\
&\times& exp{\ i<\boldsymbol{\pi}\cdot\dot{\mathbf{z}}+\boldsymbol{\pi}^*\cdot\dot{\mathbf{z}}^*-H_{CP}>} \  .
\end{eqnarray}
To compare we modify the expression (\ref{Zsigma}) introducing two  auxiliary variables $s$ and $\pi_s$, 
\begin{equation}
\label{ZsigmaMod}
Z_\sigma=\int\mathscr{D}\vec{n}\mathscr{D}\vec{\pi}\mathscr{D}\vec{\pi}_s\mathscr{D}\pi_s\delta(|\vec{n}|^2-1)
\delta(\vec{n}\cdot\pi)\delta(s)\delta(\pi_s)
e^{<\vec{\pi}\cdot\dot{\vec{n}}-H_\sigma>}\ ,
\end{equation}
and perform the change of variables $M$
$(\vec{n},\vec{\pi},s,\pi_s)\leftrightarrow(\mathbf{z},\boldsymbol{\pi})$ defined by,
\begin{eqnarray}\label{ch}
n_i&=&z^*_a\sigma_{iab}z_b\quad , \quad
\pi_i=\frac{\pi_a\sigma_{iab}z_b+z_a^*\sigma_{iab}\pi^*_b}{2}  \\
s&=&\frac{\nabla^2z^*_az_a-z^*_a\nabla^2z_a}{2i}\quad , \quad
\pi_s=\frac{z_a\pi_a-z^*_a\pi^*_a}{2i}\ .
\end{eqnarray}
The $\delta$ functions in (\ref{ZsigmaMod}) map onto  $\delta$ functions of the partition function of the $CP_1$ model  (\ref{ZCP}) 
and  the Hamiltonian actions  map into each other. The Jacobian of the transformation is $J=\det M$ with
\begin{widetext}
\begin{equation}
M=  \left(\begin{array}{cccccccc}
 z_2^* &  z_1^* &   z_2 &   z_1 & 0 &  0 &  0 &  0 \\
iz_2^* &  -iz_1^* &  -iz_2 &  iz_1 &  0 &  0 &  0 &  0 \\
 z_1^* &  -z_2^* &  z_1 &  -z_2 &  0 &  0 &  0 &  0 \\
\  [\nabla^2,z_1^*]/2i  & [\nabla^2,z_2^*]/2i &
[z_1,\nabla^2]/2i & [z_2,\nabla^2]/2i &
 0 &  0 &  0 &  0 \\
\pi_2/2 & \pi_1/2 & \pi_2^*/2 & \pi^*_1/2 &
 z_2/2 &  z_1/2 &  z_2^*/2 &  z_1^*/2 \\
 i\pi_2/2 &  -i\pi_1/2 & -i\pi_2^*/2 &  i\pi^*_1/2 &
 -iz_2/2 &  iz_1/2 &  iz_2^*/2 &  -iz_1^*/2 \\
\pi_1/2 &  -\pi_2/2 & \pi_1^*/2 &  -\pi^*_2/2 &
 z_1/2 &  -z_2/2 &  z_1^*/2 &  -z_2^*/2 \\
\pi_1/2i & \pi_2/2i &  -\pi_1^*/2i &  -\pi^*_2/2i &
 z_1/2i &  z_2/2i &  -z_1^*/2i &  -z_2^*/2i
\end{array}\right)\delta^2(\vec{x}-\vec{y})\nonumber .
\end{equation}
Using the block structure of $M$ we have
\begin{eqnarray}\scriptstyle
|J|&=&\frac{1}{2i}\det\left(\begin{array}{cccc}
 z_2^* &  z_1^* &  z_2 &  z_1 \\
 iz_2^* &  -iz_1^* &  -iz_2 &  iz_1 \\
 z_1^* &  -z_2^* &  z_1 &  -z_2  \\
\nabla^2z_1^*-z_1^*\nabla^2 & \nabla^2z_2^*-z_2^*\nabla^2 &
 z_1\nabla^2-\nabla^2z_1 &  z_2\nabla^2-\nabla^2z_2
\end{array}\right)
\times\frac{1}{2i}\left[\frac{1}{2}\right]^3
\det\left(\begin{array}{cccc}
 z_2 &  z_1 &  z_2^* &  z_1^* \\
 -iz_2 &  iz_1 &  iz_2^* & -iz_1^* \\
 z_1 & -z_2 &  z_1^* & -z_2^* \\
 z_1 &  z_2 & -z_1^* & -z_2^*
\end{array}\right).\nonumber\end{eqnarray}
\end{widetext}
Using  the identity
$(\nabla^2z^*_a)z_a+z^*_a(\nabla^2z_a)=-2|\partial_iz|^2$ we finally obtain
\begin{eqnarray}
|J|&=&\Big{|}\det\left[-\frac{1}{8.(4)}(-4i).4i
(|\partial_iz|^2+\nabla^2)\right]\Big{|}\nonumber \\
&=&\Big{|}\det\left(\frac{1}{2}[|\partial_iz|^2+\nabla^2]\right)\Big{|}\ ,
\end{eqnarray}
which is the Fadeev Popov determinant in (\ref{ZCP}).  This establishes the quantum equivalence of the theories in the sector of 
zero topological charge. The identification of the topological charges  which is preserved in the quantum theory by the canonical map, 
guarantees the quantum equivalence of the models in all the sectors.

\section{Soliton operators}
Topological solitons in field theory models are the signature of a non trivial phase structure of the quantum theory, with the phase 
transition being driven by the condensation of the quantum solitons.  Accordingly, soliton operators may be constructed in quantum field 
theory \cite{MarES1980,MarE1988,MarE1994} as a generalization of disorder operators in statistical mechanics \cite{KadLC1971}. To 
interpolate between sectors of different topological charge soliton operators should apply to the field variables the relevant topological 
behavior of the soliton solutions. For two dimensional models these ideas allow to recover Mandelstam's operator \cite{ManS1975} and the 
standard results abelian  {\cite{ColS1975} and non-abelian bosonization \cite{WitE1984,LupJR1996,ResAS2000}. They  may also be applied 
to non-abelian gauge fields \cite{MarES1989} and to fermionic 
currents \cite{MarES1992}. In 3D, soliton operators of abelian gauge theories have been investigated along this lines 
\cite{MarE1988,FurKM1990,MarEMR1992,MarE1994}. Some applications of  the $CP_1$ model Skyrmions are discussed in 
Ref. \cite{MarE2000,ConCM2010}. Here we use the canonical mapping  to construct the $\sigma-O(3)$ disorder operator from the $CP_1$ 
operator.

The topological properties  of the  $CP_1$  Skyrmion  are encoded in the   behavior in  space, like infinity and at its center given by 
\cite{RajR1982},
\begin{eqnarray}
\mathbf{z}(\vec{x}) {_{\overrightarrow {\rho \to \infty}}}
\left(\begin{array}{c}
e^{-i\arg(\vec{x})/2} \\
0
\end{array}\right)
\quad \mathbf{z}(\vec{x}){_{\overrightarrow {\rho \to 0}}}
\left(\begin{array}{c}
0 \\
e^{i\arg(\vec{x})/2}
\end{array}\right)\ , \nonumber\\
A_i(\vec{x}) {_{\overrightarrow {\rho \to \infty}}}
\frac{1}{2}\partial_i[\arg(\vec{x})]\quad A_i(\vec{x}) {_{\overrightarrow {\rho
\to 0}}}-\frac{1}{2}\partial_i[\arg(\vec{x})]\  \nonumber,
\end{eqnarray}
where $\rho$ is the radial variable and $\theta(\vec{x})=\arctan(x_1/x_2)\equiv\arg(\vec{x})$. To apply the asymptotic behavior to 
the fields, the disorder operator $\mu(x)$ should satisfy the order disorder algebra
\begin{widetext} 
\begin{eqnarray}\label{descpa}
\mu(x;c)Z_1(y)=\left\{\begin{array}{cc}
e^{-\frac{1}{2}i\arg(\vec{y}-\vec{x})}Z_1(y)\mu(x;c) & \vec{y}-\vec{x}\notin T(c)\\
Z_1(y)\mu(x;c)  & \vec{y}-\vec{x}\in T(c)
\end{array}\right.\ , \nonumber\\
\mu(x;c)Z_2(y)=\left\{\begin{array}{cc}
Z_1(y)\mu(x;c) & \vec{y}-\vec{x}\notin T(c)\\
e^{\frac{1}{2}i\arg(\vec{y}-\vec{x})}Z_1(y)\mu(x;c)  & \vec{y}-\vec{x}\in
T(c)
\end{array}\right.\ ,\nonumber \\
\label{descpb} \mu(x;c)A_i(y)=\left\{\begin{array}{cc}
\left[A_i(y)+\frac{1}{2}\partial_i\arg(\vec{y}-\vec{x})\right]\mu(x;c) & \vec{y}-\vec{x}\notin T(c)\\
\left[A_i(y)-\frac{1}{2}\partial_i\arg(\vec{y}-\vec{x})\right]\mu(x;c)  &
\vec{y}-\vec{x}\in T(c)
\end{array}\right.\ ,\nonumber
\end{eqnarray}
were $T(c)$ is a spatial region centered in $\vec{x}$ whose boundary is a plane curve $c$. 
 $\mu(x,c)$ is identified to be
\begin{eqnarray}\label{des}
\mu(\vec{x},t;c)&=&\exp\Big\{\frac{1}{2}
\int_{\mathbf{R}^2-T_{\vec{x}}}\big[Z_1(\vec{w},t)\Pi_1(\vec{w},t)-
\Pi_1^\dag(\vec{w},t)Z^\dag_1(\vec{w},t)\big]\arg(\vec{w}-\vec{x})d^2\vec{w}\nonumber \\
&-&\frac{1}{2}
\int_{T_{\vec{x}}}\big[Z_2(\vec{w},t)\Pi_2(\vec{w},t)-
\Pi_2^\dag(\vec{w},t)Z^\dag_2(\vec{w},t)\big]\arg(\vec{w}-\vec{x})d^2\vec{w}\Big\}.
\end{eqnarray}
\end{widetext}
For the $O(3)-\sigma$ model, the direct construction of the disorder variable is more complicated to pursue since the topological 
properties of the solution depend on the whole space time configuration. This is overcome by  using the canonical map. In components 
the map ($\vec{n}=\mathbf{z}^\dag\vec{\sigma}\mathbf{z}$)  is 
\begin{equation}
n_1=2\RE(z^*_1z_2)\quad n_2=2\IM(z^*_1z_2)
\quad n_3=|z_1|^2-|z_2|^2 .
\end{equation}
Using also that $\mu_\sigma^\dag=\mu_\sigma^{-1}$, it is shown that $[\mu(\vec{x}),N_3(\vec{y}]=0$  and the order disorder algebra 
(\ref{descpa}) which is non trivial only for $N_1$ and $N_2$  is  written as,
\begin{eqnarray}
\label{aodsigma}
\mu_\sigma(\vec{x})N_1(\vec{y})=\cos(\arg(\vec{y}-\vec{x}))N_1(\vec{y})\mu_\sigma(\vec{x})\ ,\nonumber\\
\mu_\sigma(\vec{x})N_2(\vec{y})=\sin(\arg(\vec{y}-\vec{x}))N_2(\vec{y})\mu_\sigma(\vec{x})\ .
\end{eqnarray}
It  does not depend on $T(c)$. The inverse of the change variables (\ref{ch}) is
\begin{eqnarray}\label{inverse}&&|z_1|=\sqrt{\frac{1+n_3}{2}}\
,\ |z_2|=\sqrt{\frac{1-n_3}{2}}\ ,\ e^{i\phi}=\frac{n_1+in_2}{\sqrt{1-n_3^2}}\ ,\nonumber\\
&&z_1=|z_1|e^{i\varphi_1}\quad,\quad z_2=|z_2|e^{i\varphi_2}
\quad,\quad\phi=\varphi_2-\varphi_1\ ,\nonumber \\
&&\pi^{CP}_1=\frac{1}{z_1}\left[\pi^\sigma_3+
\frac{i}{2}(n_1\pi^\sigma_2-n_2\pi^\sigma_1)\right]\ ,\\
&&\pi^{CP}_2=\frac{1}{z_2}\left[-\pi^\sigma_3-
\frac{i}{2}(n_1\pi^\sigma_2-n_2\pi^\sigma_1)\right]\ .\nonumber
\end{eqnarray} 
Substituting this in the classical expression of (\ref{des}) and taking into account the ordering issues for the quantum operators 
already discussed we end up with

\begin{eqnarray}\label{dess}
\mu_\sigma(\vec{x},t)&=&\exp\Big\{{i}
\int_{\mathbf{R}^2}\big[N_1(\vec{y},t)\Pi_2(\vec{y},t)-
\Pi_1(\vec{y},t)N_2(\vec{y},t)\big]\nonumber\\
&\times& \arg(\vec{y}-\vec{x})d^2y\Big\} \ ,
\end{eqnarray}
which again does not depend on $T(c)$. This operator satisfies the order disorder algebra (\ref{aodsigma}).

\section{Conclusion}
In this paper we use the  complete canonical map between the Hamiltonians descriptions which results from applying Dirac's method 
\cite{BanR1994}, of the $O(3)-\sigma$ model and the $CP_1$ model in $3D$  and show that the quantum theory is free of ordering 
ambiguities. We  demonstrate, by  exhibiting the explicit functional change of variables for the path integral and computing the 
Jacobian determinant, that the phase space partition functions of the models  computed using the complete Senjanovic's construction, 
are identical, as expected. Finally, we apply the results of the canonical equivalence to construct the $O(3)-\sigma$ soliton disorder 
operator  starting from the corresponding operator of the $CP_1$ model and verify that it satisfies the defining order disorder algebra.

\begin{acknowledgments}
This work was supported by DID-USB GID-30. 
\end{acknowledgments}

\bibliographystyle{unsrt}

\end{document}